# Type-I hyperbolic metasurfaces for highly-squeezed designer polaritons with negative group velocity

Yihao Yang[1,2,3,4], Pengfei Qin[2], Xiao Lin[3], Erping Li[2], Zuojia Wang[5,*], Baile Zhang[3,4,*], and Hongsheng Chen[1,2,*]

[1]State Key Laboratory of Modern Optical Instrumentation, College of Information Science and Electronic Engineering, Zhejiang University, Hangzhou 310027, China.

[2]Key Lab. of Advanced Micro/Nano Electronic Devices & Smart Systems of Zhejiang, The Electromagnetics Academy at Zhejiang University, Zhejiang University, Hangzhou 310027, China.

[3]Division of Physics and Applied Physics, School of Physical and Mathematical Sciences, Nanyang Technological University, 21 Nanyang Link, Singapore 637371, Singapore.

[4]Centre for Disruptive Photonic Technologies, The Photonics Institute, Nanyang Technological University, 50 Nanyang Avenue, Singapore 639798, Singapore.

[5]School of Information Science and Engineering, Shandong University, Qingdao 266237, China

*z.wang@sdu.edu.cn (Z.W.); blzhang@ntu.edu.sg (B.Z.); hansomchen@zju.edu.cn (H.C.)

## Abstract

Hyperbolic polaritons in van der Waals materials and metamaterial heterostructures provide unprecedented control over light-matter interaction at the extreme nanoscale. Here, we propose a concept of type-I hyperbolic metasurface supporting highly-squeezed magnetic designer polaritons, which act as magnetic analogues to hyperbolic polaritons in the hexagonal boron nitride (*h*-BN) in the first Reststrahlen band. Comparing with the natural *h*-BN, the size and spacing of the metasurface unit cell can be readily scaled up (or down), allowing for manipulating designer polaritons in frequency and in space at will. Experimental measurements display the cone-like hyperbolic dispersion in the momentum space, associating with an effective refractive index up to 60 and a group velocity down to 1/400 of the light speed in vacuum. By tailoring the proposed metasurface, we experimentally demonstrate an ultra-compact (with a footprint shrunken by 3600 times) integrated designer polariton circuit including high-transmission 90° sharp bending waveguides and waveguide splitters. The designed metasurface with a low profile, lightweight, and ease of access, can serve as an alternatively promising platform for emerging polaritonics, and may find many other potential applications, such as waveguiding, sensing, subdiffraction focusing/imaging, low-threshold Cherenkov radiation, strong magnetic transition enhancement, wireless energy transfer, and so forth.

## Introduction

Natural hyperbolic materials that support highly-confined hyperbolic polaritons have recently emerged as an innovative platform to confine the light at the extreme nanoscale [1-3]. For hyperbolic materials, the sign of in-plane and out-of-plane permittivity/permeability are opposite. Based on the signs of the out-of-plane permittivity/permeability, the hyperbolic materials can be classified into two types: $\varepsilon_{\perp}(\mu_{\perp}) < 0$ and $\varepsilon_{\parallel}(\mu_{\parallel}) > 0$, for type-I hyperbolic materials; $\varepsilon_{\perp}(\mu_{\perp}) > 0$ and $\varepsilon_{\parallel}(\mu_{\parallel}) < 0$, for type-II hyperbolic materials. Here, $\parallel$ and $\perp$ represent the in-plane and out-of-plane components of permittivity/permeability, respectively. According to the electromagnetic (EM) theory, the dispersion of the type-I hyperbolic media is a two-sheeted hyperboloid, and that of the type-II hyperbolic media is a single-sheeted hyperboloid[4].

As a representative naturally hyperbolic material, the polar dielectric material of hexagonal boron nitride (*h*-BN) supports hyperbolic phonon-polaritons at two separated Reststrahlen bands in the mid-infrared regime[3, 5-8]. Interestingly, its phonon polaritons in the lower (760~820 cm$^{-1}$) and upper (1,365~1,610 cm$^{-1}$) Reststrahlen bands show type-I and type-II hyperbolic dispersions, respectively. Especial attention has been given to the lower Reststrahlen band because the phonon-polaritons of the layered *h*-BN slab have opposite group and phase velocities in this band [8]. Besides, experimental investigations have demonstrated numerous merits of the phonon polaritons in the *h*-BN, such as high confinement, ultra-short wavelength, and low loss compared with metal-based surface plasmons and graphene plasmons, which makes it an excellent candidate for nano-photonics [9, 10]. The *h*-BN holds a promising future in applications for sub-diffraction imaging[5] (such as hyperlens), enhanced light-matter interaction, super-Planckian thermal emission, and so forth. However, the *h*-BN only works as a hyperbolic material in the narrow Reststrahlen frequency bands, beyond which no phonon-polaritons exist.

To mimic the hyperbolic phonon-polaritons of the *h*-BN in artificial materials and to engineer polaritons at will in frequency and space, here we propose a new platform, i.e., type-I hyperbolic metasurface with anisotropic magnetic responses. The metasurface consists of a single-layer coil array and is characterized by a negative/positive out-of-plane/in-plane permeability (This is qualitatively different from the conventional hyperbolic metasurfaces, where the in-plane surface plasmons [11] show a hyperbolic dispersion relation and propagate with a convergent manner [12-15]).

The type-I hyperbolic metasurface behaves as an artificial *h*-BN in many ways. For examples, the designer polaritons on the metasurface carry ultra-high momentum and ultra-large negative group velocity. These are demonstrated in our experiments, by direct imaging the near-field distribution, where we observe a cone-like dispersion in reciprocal space, associating with a remarkably high effective relative refractive index up to 60 and a slow group velocity down to *c*/400 (*c* is the speed of light in vacuum).

With so many unique features and merits of low profile, lightweight, and ease of access, the type-I hyperbolic metasurfaces have excellent potentials in applications. First, the high effective refractive index of the metasurface makes it an excellent candidate to design highly integrated waveguide circuits. On the one hand, a high effective refractive index always leads to a small footprint of the waveguide circuits. On the other hand, the designer polaritons smoothly travel through the sharp-corner waveguides and waveguide splitters with high transmissions, which can be explained by the quasi-static approximation[16-18]. With these favorable properties, we experimentally achieve a whole integrated polariton circuit with footprint shrunken by almost 3600 times, substantially exceeding the conventional waveguide circuits (typically a few times). It enables the whole polariton circuit to be under the diffraction limit ($\lambda_0$/2, where $\lambda_0$ is the free-space wavelength), which is highly pursued in polaritonics (including plasmonics). Besides, the large effective refractive index of the present metasurface could be used to design subdiffraction focusing/imaging devices [5], low-electron-velocity Cherenkov radiation emitters [19], and so forth. Second, the significant group velocity dramatically enhances the light-matter interaction and makes the metasurface extremely sensitive to the thickness and refraction index of the surroundings, thus an excellent optical sensor[10]. Third, due to the duality between electric and magnetic phenomena, the magnetic hyperbolic polariton with highly-squeezed modes could provide a new pathway for achieving strong magnetic transition enhancement[20]. Fourth, the process of EM energy transport on the metasurface is precisely the same as the well-known wireless energy transfer[21-23]. Therefore, the present metasurface may inspire novel wireless energy transfer devices. Finally, the present metasurface could be a new platform in polaritonics, with arbitrary dispersions in frequency and space.

**Results**

The proposed type-I hyperbolic metasurface is shown in Fig. 1(a), which consists of arrays of coiling copper wires printed on a dielectric substrate. The inset of the figure shows the details of the coils, where the golden and brown regions indicate coppers layer and dielectric substrates, respectively. Here, $a$=2 mm, $w$=$g$=0.2 mm, numbers of turns $N$=15, $p$=13.8 mm, and $t$=1 mm. The thickness of metal layer is 0.035 mm; the conductivity of copper is $5.7\times10^7$ S/m; the relative permittivity of the substrate is 2.55+0.001i below 10.0 GHz. The substrate, which is used to support the copper film and can be removed, has a slight impact on the EM properties of the metasurface. Therefore, the thickness of the metasurface can be extremely thin, e.g., $1/10^5$ times of the operational wavelength.

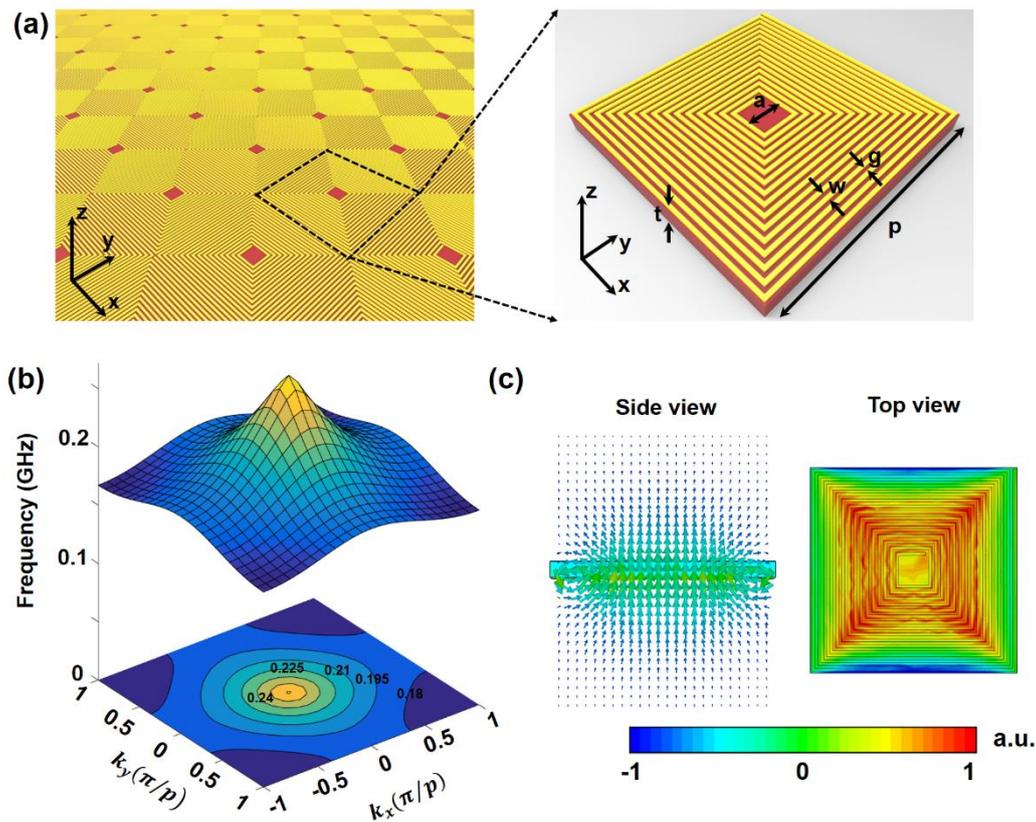

**Figure 1. Type-I hyperbolic metasurface.** (a) The proposed type-I hyperbolic metasurface, which consists of coiling copper wires printed on a dielectric substrate. The inset shows the details of the unit cell. Here, $a$=2 mm, $w$=$g$=0.2 mm, numbers of turns $N$=15, $p$=13.8 mm, and $t$=1 mm. The thickness of metal layer is 0.035 mm; the conductivity of copper is $5.7\times10^7$ S/m; the relative permittivity of the substrate is 2.55+0.001i at 10 GHz. (b) Equal-frequency contour of the dispersion of the type-I hyperbolic metasurface in the first Brillouin zone. (c) Side

view and top view of the magnetic field distribution at the magnetic resonance. The color bar measures the intensity of the magnetic field.

By employing the eigenvalue module of the commercial software Computer Simulation Technology (CST) Microwave Studio, we obtain equal-frequency contours (EFCs) of the fundamental band in the first Brillouin zone (FBZ) of the proposed metasurface (Fig. 1(b)). It is interesting that the dispersion of the fundamental band shows a topology of a cone, rather than an inverted cone in the usual cases. As the group velocity is calculated with [24]

$$v_g = \frac{\partial \omega}{\partial k}, \quad (1)$$

where $\omega$ and $k$ are the angular frequency and wave vector, respectively, the group velocities of the designer polaritons on the metasurface are negative. The eigenmodes on the metasurface are shown in Fig. 1(c), where the designer polaritons are highly confined vertically and horizontally. The magnetic field distribution indicates a strong magnetic dipole, which physically arises from the surface currents flowing along the spiral coil.

To understand the exotic behavior of the designer polaritons on the present metasurface, we construct a layered metamaterial by stacking the metasurface periodically along the $z$-direction, as shown in Fig. 2(a). As the metal coils produce $z$-oriented magnetic resonances, this artificial material works as a type-I magnetic hyperbolic metamaterial. By applying a well-established retrieval process [25], we obtained the effective constitutive parameters of the constructed metamaterial (Fig. 2(b)), where $\mu_z$ is negative from 0.315 GHz to 0.4 GHz, while the other constitutive parameters are positive. Then we study the EM properties of a metamaterial slab with a finite thickness in the $z$-direction. Considering the fundamental transverse electric (TE) even mode (the inset of Fig. 2(c)), the corresponding dispersion relation of the designer polaritons propagating in the metamaterial slab is

$$k_d - \frac{k_m}{\mu_y}\tanh(k_m d/2) = 0,$$

$$\text{with} \quad k_d = \sqrt{\beta^2 - k_0^2}, \, k_m = \sqrt{(\beta^2/\mu_z - k_0^2 \varepsilon_x)\mu_y}, \quad (2)$$

where $d$ is the thickness of the metamaterial slab, $\beta$ represents the wavevector of the designer polaritons along the propagating direction, and $k_0$ denotes the wavevector in vacuum. By substituting the retrieval constitutive parameters to Eq. (2), one can obtain the dispersion relation of the metamaterial slab with different thicknesses. Here, we consider the metamaterial with 1, 2, 3, and 4 layers. By choosing the proper effective thicknesses, we find that the calculated dispersions match with the simulated counterparts excellently. Note that as the number of layers is small, the edge effects make the EM properties of the metamaterial slightly change. Therefore we use the effective thicknesses rather than the physical thicknesses. Besides, we visualize the dispersion of the surface plasmons via a false-color plot of |Im (reflection)|, where reflection is the reflection coefficient of TE wave [26]. This is because the designer polaritons propagating in the metamaterial slab are the singularity poles in the reflection coefficient. Interestingly, Eq. (2) holds true even the metamaterial slab with only a single layer, namely a metasurface, which manifests that the z-oriented negative magnetic response mainly arises from the self-induced inductances and capacitances inside the coils rather than the interlayer coupling.

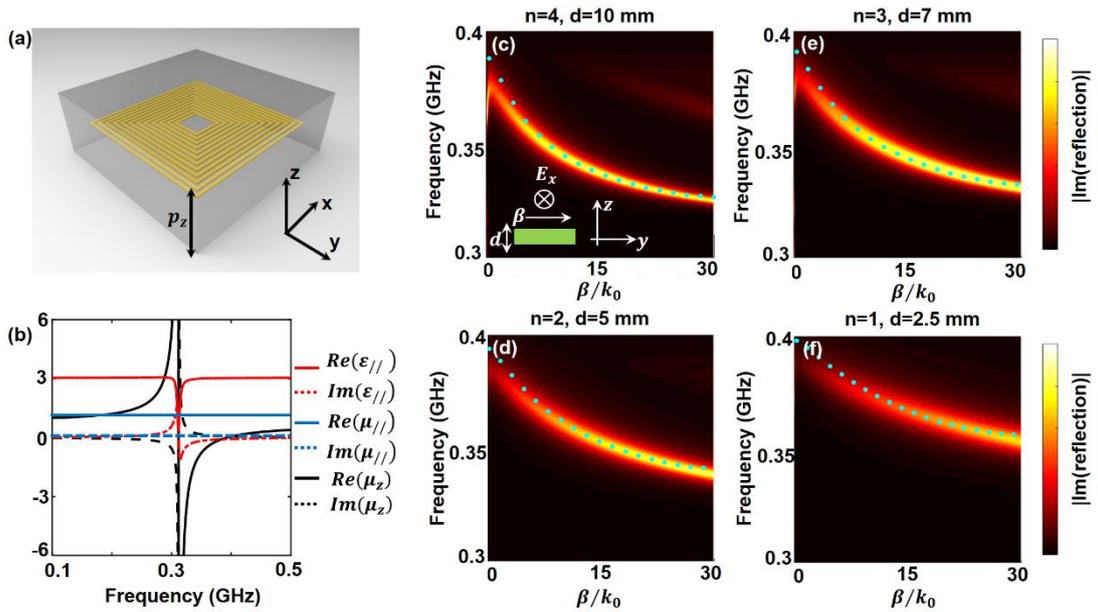

**Figure 2. From type-I hyperbolic metamaterial to metasurface.** (a) Scheme of the hyperbolic metamaterial composed of coiling copper wires. Here, $p_z$=5 mm, $n$=12, $p$=13.8 mm, $w$=$g$=0.2 mm, and $a$=2 mm. The coiling metal wires are embedded in a dielectric host with a relative permittivity of 2.55. (b) Retrieved constitutive parameters of the hyperbolic metamaterial. The hyperbolic region is from 0.315 GHz to 0.4 GHz. $\varepsilon_{//}$, $\mu_{//}$ and $\mu_z$ are the in-plane relative permittivity and permeability, and out-of-plane relative permeability, respectively. (c)-(f) Dispersions of the hyperbolic metamaterial slab with 4, 3, 2, and 1 periods along the z-direction. The green dots

are the dispersion relation of the polaritons propagating in the slab with practical metamaterial structures. The false-color plots are the dispersion relations of the polaritons propagating in a homogeneous slab with the retrieval constitutive parameters. Here, the *d* is the effective thickness of the metamaterial slab.

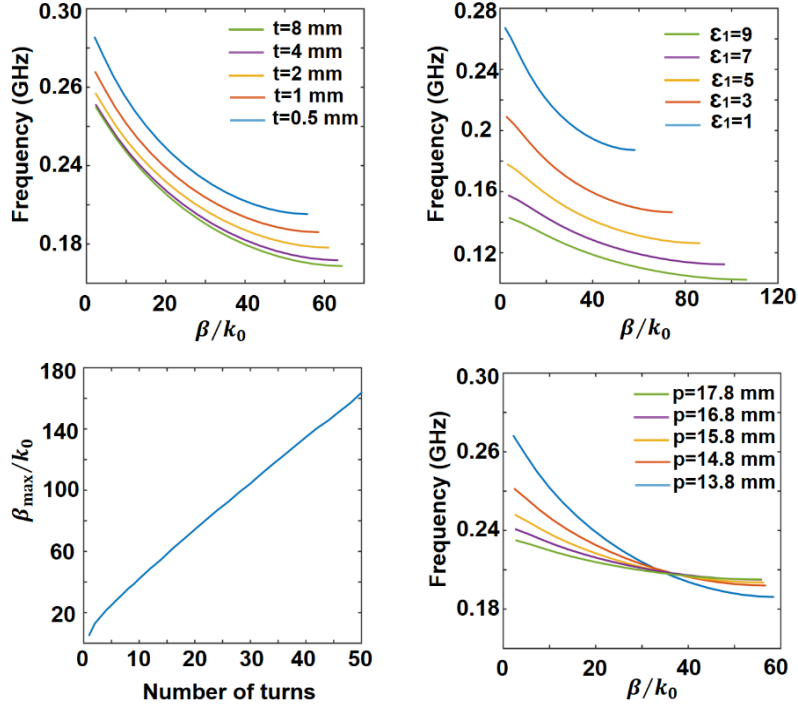

**Figure 3. Dependence of the designer hyperbolic polaritons on lattice structure.** (a) Dispersions of the metasurface with different thicknesses of the substrate. (b) Dispersions of the metasurface with different permittivity of the substrate. (c) Maximum wave vector as a function of numbers of turns. (d) Dispersions of the metasurface with different periods of the unit cell, where the coils keep the same.

In the following, we study the impacts of the geometries and the permittivity changes on the behavior of the designer polaritons on the metasurface. When altering the thicknesses of the dielectric substrate below 4 mm (about 1/750 of the operational space wavelength), the dispersions dramatically change, which manifests that the EM responses of the metasurface are very sensitive to the thicknesses (Fig. 3(a)). When the thickness is over 4 mm, the dispersions almost stay the same, indicating that the designer polaritons are highly confined around the metasurface and dramatically decay into the background (Fig. 3(a)). Besides, the dispersion of the metasurface is also very sensitive to the permittivity of the substrate (Fig. 3(b)). Such a high sensitivity to the thicknesses and permittivity of the substrates origins from the strong light-matter interaction in the metal coils, and may find applications in optical sensing [27, 28].

We also study the relations between the number of turns and the maximum of the squeezing factor, here, the squeezing factor or the effective relative refractive index is defined as $\beta/k_0$. One can see that the maximum of the squeezing factor is almost proportional to the number of turns (Fig. 3(c)). In 2D materials, a larger squeezing factor in 2D materials always associates with a larger group velocity, and thus a severe propagation loss, which imposes a limitation for the largest squeezing factor [10]. However, for the present metasurface, removing the lossy substrate and using the superconductor coils [29], the ohmic loss is almost neglectable, and therefore there is no limitation for the largest squeezing factor. Finally, we change the period and keep the other parameters the same. One can see that when enlarging the period, the bandwidths of the dispersions decreases and the group velocity increases (Fig. 3(d)). This is because the coupling between two neighbor unit cells decreases when increasing the period, and the energy can only transmit via the strong magnetic resonance of the coils at the frequency range centering on the magnetic resonance of a single coil. The limiting case is that the period is so large that the dispersion becomes almost a flat line which is almost at the magnetic resonance frequency of a single isolated coil. In that case, the physical mechanism of the energy transfer between two neighbor coils is identical to the so-called wireless energy transfer [22].

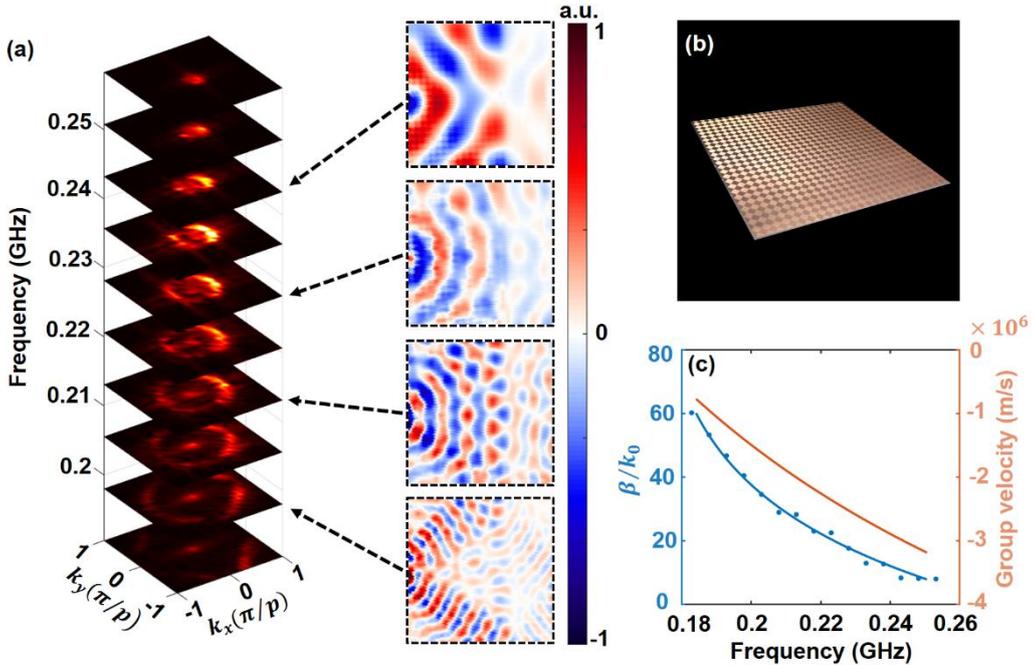

**Figure 4. Measured field distributions, dispersions, and group velocities of the designer polaritons on the type-I hyperbolic metasurface.** (a) Measured dispersion relation of the surface plasmons on the metasurface in the FBZ, and $H_z$ field distributions in the plane 2 mm over the metasurface at 0.198 GHz, 0.213 GHz, 0.228 GHz, and 0.243 GHz, respectively. (b) Photograph of the metasurface sample. (c) Measured normalized wave vectors and group velocities of the surface plasmons at different frequencies. Here, the blue dots are the experimental data; the blue curve represents exponential fits to the data points of wave vectors; the yellow curve is the group velocity obtained from the blue curve.

Several experiments are carried out to characterize the proposed metasurface. The photograph of the fabricated sample is shown in Fig. 4(b). In experiments, a port directly connects to one of the coil unit cells at the side of the metasurface, to excite the designer polaritons. The receiver antenna is a compact coil antenna with a magnetic resonance around 0.2 GHz. The receiver antenna is fixed at the arm of a three-dimensional movement platform. Both of the antennas connect to the vector network analyzer (VNA) to get the amplitude and phase of the magnetic field. With the above system, we scan the $z$-oriented magnetic field distributions in the plane 5 mm above the metasurface, as shown in the right column of Fig. 4(a). Then we apply the spatial Fourier transform to obtain the momentum space of the designer polaritons (Fig. 4(a)).

Our measured results show a cone in the momentum space, which verifies the theoretical prediction. Counterintuitively, we directly observe that the wavelength becomes longer as the frequency increases. We also retrieve the squeezing factor of the designer polaritons from the experimental data. Impressively, the squeezed factor ranges from 8 up to 60, which breaks the limitation of the ordinary metamaterials and metasurfaces [2]. Such a large squeezing factor is comparable or even larger than that in the two-dimensional materials, such as graphene plasmons [10]. We should note that with more turns of the coils, the squeezing factor can be even larger and has no strict limits. Such a high squeezing factor is highly pursued in miniaturization of the integrated waveguide circuits. Here, the squeezing factor is as large as 60, which means we can use it to shrink the footprint of integrated waveguide circuits by almost 3600 times, substantially exceed the conventional waveguide circuits (typically a few times). Besides, we also retrieve the group velocity of the designer polaritons from the fitted dispersion by applying Eq. (1), which ranges from $3\times10^6$ m/s to $7.5\times10^5$ m/s, or from 0.01 $c$ to 0.0025 $c$. Therefore, such a metasurface shows an ultra-slow light effect in a relative bandwidth of 32.6%. The broadband ultraslow light effect of the metasurface may lead to numerical applications, such as delay lines, EM energy storage, and strongly enhanced light-matter interaction.

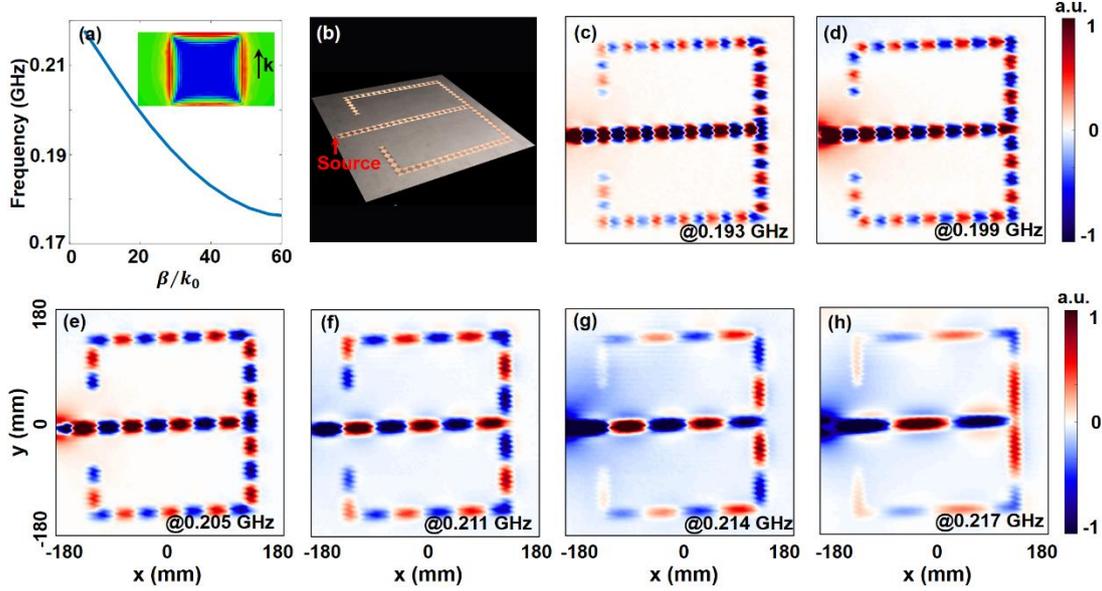

**Figure 5. A deep-sub-wavelength integrated waveguide circuit based on the type-I hyperbolic metasurface.** (a) Dispersion of designer polaritons on the meta-ribbon of type-I hyperbolic metasurfaces. (b) A fabricated sample of a waveguide circuit based on the type-I hyperbolic metasurface. The red arrow represents the location of the source. (c)-(h) Measured $H_z$ field distributions in the plane 2 mm over the metasurface at 0.193 GHz (c), 0.199 GHz (d), 0.205 GHz (e), 0.211 GHz (f), 0.211 GHz (g), and 0.217 GHz (h), respectively. Here the whole measured region is below the diffraction limit, i.e., half of the wavelength in free space.

As the type-I hyperbolic metasurface shows so many fascinating properties, it could find plenty of applications. As an example, by taking advantages of the high squeezing factor of the type-I hyperbolic metasurface, we design an ultra-compact integrated waveguide circuit with the ultra-small footprint. We first tailor the metasurface to meta-ribbons with a single-unit-cell width. The tailored meta-ribbons still maintain the high squeezing factor of the metasurface (Fig. 5(a)). With the meta-ribbons, we construct a waveguide splitter with several 90° sharply twisted corners (Fig. 5(b)). From the measured magnetic field distribution over the waveguide splitter (Figs. 5(c)-(h)), one can see that the designer polaritons are launched at the excitation, split into two beams at the splitter, and smoothly pass through the 90° sharp corners of the waveguide.

Interestingly, the transmission of designer polaritons through the splitter and sharp corners are very high. This is because the effective width of the meta-ribbon is much smaller than the effective wavelength of the designer polaritons, and we can apply a common quasi-static approximation [16-18]. With this approximations, we can consider the sharp-corner waveguide as a junction with two transmission lines with the same impedance and the splitter as a junction with one input transmission line and two output transmission lines with the same impedance. Therefore, the reflection loss of the bending waveguide can be neglectable, and that of the splitter is about 12%. Note that such a high transmission can also be found in photonic crystal waveguides [30, 31] and the plasmonic waveguides [16-18]. Besides, one can see that the meta-ribbons work excellently from 0.193 GHz to 0.217 GHz. We should emphasize here that the whole structure is only with a size of 1/4 free-space wavelength. Therefore, we experimentally achieve an ultra-compact integrated designer polariton circuit beyond the diffraction limit.

**Conclusions**

In summary, our work identifies a new class of hyperbolic metasurface, namely type-I hyperbolic metasurface, which behaves in many ways the same as an artificial $h$-BN in the first Reststrahlen band, such as an extremely high effective refractive index, and an ultra-large and negative group velocity. Compared with the natural $h$-BN, the artificial type-I hyperbolic metasurface is readily geometry-tailorable, allowing for the creation of designer polaritons with almost arbitrary dispersions in both frequency and space. The present metasurface with a low profile, lightweight, and ease of access, could serve as a new platform in polaritonics and may find many other potential applications, such as sensing, subdiffraction focusing/imaging, low-electron-velocity Cherenkov radiation emission, strong magnetic transition enhancement, and wireless energy transfer. In combination with flexible substrates[32] and active and nonlinear components[33, 34], we envision further exciting possibilities such as actively controlling the conformal designer polaritons with nonlinear properties at deeply subwavelength scale. Although demonstrated in microwave frequency, the concept of type-I hyperbolic metasurface shows the generality and could apply to the other frequencies.

**Acknowledgments**

The work was sponsored by the National Natural Science Foundation of China under Grant Nos.

61625502, 61574127, 61601408, 61775193 and 11704332, the ZJNSF under Grant No. LY17F010008, the Top-Notch Young Talents Program of China, the Fundamental Research Funds for the Central Universities under Grant No. 2017XZZX008-06, and the Innovation Joint Research Center for Cyber-Physical-Society System.

## Authors Contributions

Y.Y. conceived the original idea and designed the metasurface. Y.Y. and P.Q. carried out the experiments. Y.Y., Z.W., X.L., E.L., B.Z., and H.C. produced the manuscript and interpreted the results. Z.W., and B.Z., and H.C. supervised the project. All authors participated in discussions and reviewed the manuscript.

## Competing Financial Interests

The authors declare no competing financial interests.

## Data availability

The data that support the plots within this paper and other findings of this study are available from the corresponding author upon request